\documentclass[pra,twocolumn,showpacs,preprintnumbers,amsmath,amssymb]{revtex4}
\usepackage{graphicx}

\newtheorem{theorem}{Theorem}
\newtheorem{lemma}{Lemma}
\newtheorem{remark}{Remark}
\newtheorem{example}{Example}

\newcommand{\enlem}{\hfill $\Diamond$ \end{lemma}}
\newcommand{\closedef}{\hfill $\Diamond$ \end{definition}}
\newcommand{\enth}{\hfill $\Diamond$ \end{theorem}}
\newcommand{\encor}{\hfill $\Diamond$ \end{corollary}}
\newcommand{\enprop}{\hfill $\Diamond$ \end{proposition}}
\newcommand{\encond}{\hfill $\Diamond$ \end{condition}}
\newcommand{\exam}[1]{\begin{example}\label{ex:#1}}
\newcommand{\enexam}{\QED\end{example}}
\newcommand{\beremark}[1]{\begin{remark}\label{rmk:#1}}
\newcommand{\enremark}{\QED\end{remark}}

\newcommand{\mymathbb}[1]{{\mathbb #1}} 

\newcommand{\cA}{{\cal A}}
\newcommand{\sA}{\mathsf{A}}

\newcommand{\sB}{\mathsf{B}}

\newcommand{\cC}{{\cal C}}

\newcommand{\cD}{{\cal D}}

\newcommand{\sF}{\mathsf{F}}
\newcommand{\myF}{{\sF}}

\newcommand{\sH}{\mathsf{H}}
\newcommand{\cI}{{\cal I}}

\newcommand{\sL}{\mathsf{L}}
\newcommand{\cM}{{\cal M}}

\newcommand{\sN}{\mathsf{N}}

\newcommand{\sP}{\mathsf{P}}
\newcommand{\cQ}{{\cal Q}}

\newcommand{\cR}{{\cal R}}

\newcommand{\cT}{{\cal T}}

\newcommand{\cX}{{\cal X}}

\newcommand{\bZ}{\mymathbb{Z}}

\newcommand{\vep}{\varepsilon}
\renewcommand{\phi}{\varphi}
\renewcommand{\subset}{\subseteq}

\newcommand{\mbm}[1]{#1}

\newcommand{\cmple}{^{\rm c}}

\newcommand{\field}{\mymathbb{F}}
\newcommand{\Capa}{\mathsf{C}}
\newcommand{\Erc}{E_{\rm r}}
\newcommand{\Pe}{{\rm P_e}}
\newcommand{\Pstar}[2]{{\rm P}^{\star}_{#1,#2}}

\newcommand{\Wch}{W}
\newcommand{\Vch}{V}

\newcommand{\tnsr}{\otimes}

\newcommand{\lag}{\langle}
\newcommand{\rag}{\rangle}
\newcommand{\crd}[1]{|#1|}
\newcommand{\bra}[1]{\lag #1 |}
\newcommand{\ket}[1]{| #1 \rag}
\newcommand{\indc}{{\bf 1}}
\newcommand{\syp}[2]{( #1,  #2 )_{\rm sp}}
\newcommand{\dmn}{d}
\newcommand{\Hch}{{\sH}}
\newcommand{\Hgn}{{\sH}}
\newcommand{\Bop}{\sL}
\newcommand{\Hcd}{\cC}

\newcommand{\Ebasis}{\sN}
\newcommand{\Ebe}{N}

\newcommand{\Icr}{J} 
\newcommand{\Fbar}{\overline{F}}
\newcommand{\Aso}{\sA}  
\newcommand{\Acn}{\sA}  
\newcommand{\Bcn}{\sB}  
\newcommand{\CPex}{\cM}
\newcommand{\CPexO}{M}
\newcommand{\Cso}{L}

\newcommand{\Pht}{\widehat{P}}

\newcommand{\tnsn}{^{\tnsr n}}

\begin{document} 

\title{
Exponential lower bound on the highest fidelity achievable by\\
quantum error-correcting codes}

\author{Mitsuru Hamada}
\email{mitsuru@ieee.org}

\affiliation{Quantum Computation and Information Project (ERATO)\\
     Japan Science and Technology Corporation,
      5-28-3, Hongo, Bunkyo-ku, Tokyo 113-0033, Japan
}

\date{Sep.\ 21, 2001; 
 Received, Phys.\ Rev.: Oct.\ 12, 2001; Published, Phys.\ Rev.\ A, {\bf 65}, 052305, Apr.\ 15, 2002}

\begin{abstract}
On a class of quantum channels which includes the depolarizing channel, 
the highest fidelity
of quantum error-correcting codes of length $n$ and rate $R$
is proven to be lower bounded by
$1 - \exp [-n E(R)+o(n)]$ for some function $E(R)$.
The $E(R)$ is positive below some threshold $R_0$,
which implies $R_0$ is a lower bound on the quantum capacity.
\end{abstract}


\maketitle

\section{Introduction}

Quantum error-correcting codes (simply called codes in this paper) 
are deemed 
indispensable for quantum computation as schemes that protect
quantum states from decoherence. 
An information theoretic problem relevant to such codes is one of determining
the quantum capacity of a channel, which is far from 
settled%
~\cite{shor95,schumacher96, bennett96m, dss98,barnum00}.
This paper treats a problem closely related to the quantum capacity.
The corresponding problem in classical information theory
is that of determining the the highest error exponent,
called the {\em reliability function},\/
of a channel~\cite{gallager,csiszar_koerner}, which is briefly reviewed here.
A classical memoryless channel over a finite alphabet $\cX$ is
a set of conditional probabilities $\{ \Wch(v|u) \}_{u,v\in\cX}$.
A classical code is a pair $(\cC, \cD)$ consisting of
a codeword set $\cC\subset\cX^{n}$ and
a decoding map $\cD : \cX^n \to \cC$.
The performance of a classical code is evaluated in terms of
maximum decoding error probability
\[
\Pe(\cC,\cD) = \max_{\mbm{x}\in\cC} \sum_{\mbm{y} : \ \cD(\mbm{y}) \ne \mbm{x}} 
\Wch^n(\mbm{y}|\mbm{x}),
\]
where $\Wch^{n}(y_1\dots y_n | x_1\dots x_n)=
\Wch(y_1| x_1) \dots \Wch(y_n| x_n)$.
For fixed $n$ and $k$, let 
$\Pstar{n}{k}$ denote the minimum of $\Pe(\cC,\cD)$
over all possible choices of $(\cC, \cD)$ with 
$\log_{\dmn'} \crd{\cC} \ge k$, where
the base $\dmn'>1$ is arbitrarily fixed.
Shannon's channel coding theorem states that
if a rate $R$ is less than the capacity $\Capa(\Wch)$ 
of the channel $\Wch$,  then $\Pstar{n}{Rn}\to 0$.
A stronger result 
of large-deviation theoretic appearance 
has been known%
~\cite{gallager65,gallager,csiszar_koerner,litsyn99}:
There exists a function $\Erc(R,\Wch)$, 
called the {\em random coding exponent}\/ of $\Wch$,
such that
\begin{equation}\label{eq:error_exp1}
 \liminf_{n\to\infty} - \frac{1}{n}\log_{\dmn'} 
\Pstar{n}{Rn} \ge \Erc(R,\Wch),
\end{equation}
i.e., $\Pstar{n}{Rn} \lesssim \exp_{\dmn'} [ -n \Erc(R,\Wch) ]$,
and 
\begin{equation} \label{eq:error_exp2}
\Erc(R,\Wch)>0 \quad \mbox{ if } \quad R< \Capa(\Wch). 
\end{equation}
The 
function
$E^{\star}(R,\Wch) = \liminf_{n} - \frac{1}{n}\log_{\dmn'} \Pstar{n}{Rn}$,
which is called the reliability function of $\Wch$,
shows a trade-off 
between the reliability and data transmission rate
of the best codes on channel $\Wch$. 
The $E^{\star}(R,\Wch)$ actually equals 
$\Erc(R,\Wch)$ for relatively large rates $R$, 
and complete determination of $E^{\star}(R,\Wch)$
is one of the central issues
in classical information theory%
~\cite{gallager,gallager65,csiszar_koerner,litsyn99}.
Note that Shannon's coding theorem directly follows from
(\ref{eq:error_exp1}) and (\ref{eq:error_exp2}).

Motivated by this classical issue, this paper presents
an exponential lower bound on 
the highest possible 
fidelity of a code used on a class of quantum channels,
which includes the depolarizing channel often discussed in the
literature~\cite{bennett96m,dss98,preskillLN}.
This work was inspired by the recent 
result of Matsumoto and Uyematsu~\cite{MatsumotoUyematsu01},
who used
an algebraic fact due to Calderbank et al.~\cite{crss97} 
[Eq.~(\ref{eq:C_uniform}) below] to deduce a lower bound on
the quantum capacity.
This work's approach resembles theirs in that
both bounds are shown using random coding arguments%
~\cite{gallager,csiszar_koerner,goppa74} based on (\ref{eq:C_uniform}), 
but differs
from \cite{MatsumotoUyematsu01} 
in that while \cite{MatsumotoUyematsu01} uses an analog of 
minimum Hamming distance decoding,
this work employs
an analog of minimum entropy decoding~\cite{CsiszarKoerner81a}
together with the method of types from classical 
information theory~\cite{csiszar_koerner, CsiszarKoerner81a, csiszar98}, 
which enables us to obtain the exponential bound 
analogous to (\ref{eq:error_exp1})
in a simple enumerative manner. 

\section{Exponential Bound on Fidelity}

We follow the standard formalism
of quantum information theory
which assumes 
all possible quantum operations
and state changes,
including the effects of
quantum channels, 
are described in terms of 
{\em completely positive}\/ (CP) linear maps%
~\cite{choi75,schumacher96,barnum00}.
In this paper, only 
{\em trace-preserving completely positive}\/ (TPCP) linear maps are treated.
Given a Hilbert space $\Hgn$ of finite dimension,
let $\Bop(\Hgn)$ denote the set of linear operators on $\Hgn$. 
In general, every CP linear map $\CPex: \Bop(\Hgn) \to \Bop(\Hgn)$ 
has an operator-sum representation 
$\CPex(\rho) = \sum_{i\in\cI} \CPexO_i \rho \CPexO_i^{\dagger}$ for some
$\CPexO_i\in\Bop(\Hgn)$, $i\in\cI$~\cite{choi75,schumacher96}.
When $\CPex$ is specified by a set of operators
$\{ \CPexO_i \}_{i\in\cI}$, which is not unique, in this way, we write
$\CPex \sim \{ \CPexO_i \}_{i\in\cI}$. 

Hereafter, $\Hch$ denotes an arbitrarily fixed Hilbert
space whose dimension $\dmn$ is a prime number.
A quantum channel is a sequence of
TPCP linear maps 
$\{ \cA_n  : \Bop(\Hch^{\tnsr n}) \to \Bop(\Hch^{\tnsr n}) \}$.
We want a large subspace $\Hcd=\Hcd_n \subset \Hch^{\tnsr n}$
in which every state vector remains almost unchanged 
after the effect of a channel
followed by the action of some suitable recovery process.
The recovery process is again described as a TPCP linear map
$ 
\cR_n: \Bop(\Hch^{\tnsr n}) \to \Bop(\Hch^{\tnsr n}). 
$ 
A pair $(\Hcd_n, \cR_n)$ consisting of such a subspace $\Hcd_n$
and a TPCP linear map $\cR_n$
is called a {\em code}\/
and its performance is evaluated in terms of minimum fidelity%
~\cite{KnillLaflamme97,dss98,barnum00}
\[
F(\Hcd_n, \cR_n\cA_n) = \min_{ \ket{\psi} \in \Hcd_n } 
\lag\psi| \cR_n\cA_n(|\psi\rag \lag\psi|) |\psi\rag,
\]
where $\cR_n\cA_n$ denotes the composition of $\cA_n$ and $\cR_n$.
Throughout, bras $\bra{\cdot}$ and kets $\ket{\cdot}$ are
assumed normalized.
A subspace $\Hcd_n$ alone is also called a code
assuming implicitly some recovery operator.
Let $F_{n,k}^{\star}(\cA_n)$ denote the
supremum of
$F(\Hcd_n, \cR_n\cA_n)$ such that there exists a code $(\Hcd_n, \cR_n)$
with $\log_{\dmn} \dim \Hcd_n \ge k$.
This paper gives an
exponential lower bound on $F_{n,k}^{\star}(\cA_n)$,
in the case where $\{\cA_n\}$ is a slight generalization
of the depolarizing channel 
specified as follows.

Fix an orthonormal basis 
$\{ |0\rag,\dots, |\dmn-1\rag \}$ of $\Hch$.
Put $\cX=\{0,\dots,\dmn-1\}^2$
and $\Ebe_{(i,j)} = X^i Z^j$ for $(i,j)\in \cX$,
where 
the unitary operators $X, Z \in \Bop(\Hch)$ are defined by
\begin{equation}\label{eq:error_basis}
X |j \rag  = |(j-1) \bmod \dmn \, \rag, \quad
Z |j \rag = \omega^ j |j \rag
\end{equation}
with $\omega$ being a primitive $\dmn$-th root of unity%
~\cite{knill96a,knill96b}.
The $\{ \Ebe_{u} \}_{u\in\cX}$ is a basis of $\Bop(\Hch)$
and a generalization of the Pauli operators (including the identity) 
in that when $\dmn=2$, the basis $\{ I, X, XZ, Z \}$
is the same as the set of Pauli operators
up to a phase factor.
For simplicity, we confine ourselves to treating analogs of what are called 
memoryless channels in classical information theory,
i.e., those $\{ \cA_n \}$ such that $\cA_n = \cA^{\tnsr n}$,
$n=1,2,\dots$, for some $\cA : \Bop(\Hch) \to \Bop(\Hch)$;
such a channel $\{ \cA^{\tnsr n} \}$ 
is referred to as 
the memoryless channel $\cA$.
In addition, we treat only channels that can be written as
$\cA \sim \{ \sqrt{P(u)} \Ebe_{u} \}_{u\in\cX}$,
where $P$ is a probability distribution on $\cX$.
This restriction is mainly due to that the codes to be proven
to have the desired performance are
{\em symplectic (stabilizer, or additive) codes}\/%
~\cite{crss97,gottesman96,knill96a,knill96b,rains99,AshikhminKnill00},
which exploit some 
algebraic property of the basis 
$\{ \Ebe_{u} \}_{u\in\cX}$,
and that analysis of code performance naturally turns out to
be easy for this class of channels.
Analysis for a wider class of channels will be given
in future papers.

As is usual in information theory, 
the classical informational divergence or relative entropy
is denoted by $D$ and entropy by $H$~\cite{csiszar_koerner,csiszar98}:
for probability distributions $P$ and $Q$ on a finite set $\cX$,
$D(P||Q)=\sum_{x\in\cX} P(x) \log_{\dmn} [P(x)/Q(x)]$ and
$H(Q)= - \sum_{x\in\cX} Q(x) \log_{\dmn} Q(x)$.
This paper's main result is 
\begin{theorem}\label{th:main}
Let integers $n$, $k$ and
a real number $R$ satisfy $0 \le k \le Rn $ and $0 \le R < 1$ 
(a typical choice is $k = \lfloor Rn \rfloor$ for an arbitrarily 
fixed rate $R$).
Then, for a memoryless channel 
$\cA \sim \{ \sqrt{P(u)} \Ebe_{u} \}_{u\in\cX}$, we have
\begin{equation*}
 F_{n,k}^{\star}(\cA^{\tnsr n}) \ge 1 - (n+1)^{2(\dmn^2-1)} \dmn ^{ - n E(R,P) }
\end{equation*}
where
\[
E(R,P)=\min_{Q} [ D(Q||P) + |1-H(Q)-R|^+ ],
\]
$|x|^+ =\max\{x,0\}$, and
the minimization with respect to $Q$ is taken 
over all probability distributions on $\cX$. 
\end{theorem}
%

{\em Remarks:}\/
An immediate consequence of the theorem is that
the quantum capacity~\cite{schumacher96, bennett96m, dss98,barnum00} 
of $\cA$ is lower bounded by $1-H(P)$.
To see this, observe that $E(R,P)$ is positive for $R<1-H(P)$
due to the basic inequality $D(Q||P) \ge 0$ where equality occurs
if and only if $Q=P$~\cite{csiszar_koerner}.
%
%
The bound $1-H(P)$ appeared earlier 
in \cite{preskillLN}, Sec.~7.16.2.

\begin{figure}
\includegraphics[scale=0.7]{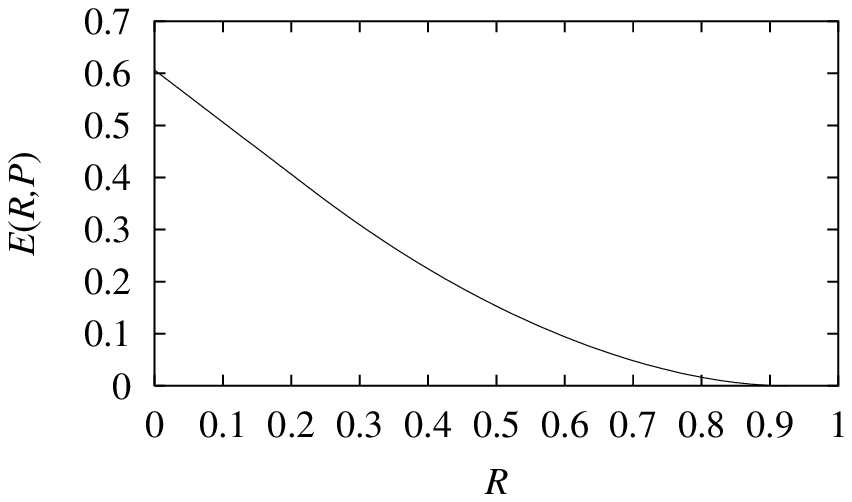} 
\caption{\label{fig:1} The function $E(R,P)$ for the depolarizing channel,
where $\dmn=2$
and $P((0,0))=1-3\vep$, $P(u)=\vep$ for $u \ne (0,0)$, 
$u \in \cX = \{ 0,1 \}^2$,
with $\vep=0.0025$.}
\end{figure}

Another direct consequence of the theorem is
\begin{equation}\label{eq:new_exp}
\liminf_{n\to\infty} - \frac{1}{n}\log_{\dmn}\{1-F^{\star}_{n,\lfloor Rn \rfloor}(\cA\tnsn)\} 
\ge E(R,P),
\end{equation}
which resembles (\ref{eq:error_exp1}).
In fact, we can see that $E(R,P)$ is closely related to 
$\Erc(R,\Wch)$ in (\ref{eq:error_exp1}) as follows.
A specific form of 
$\Erc$ 
is $\Erc(R,\Wch)= \max_{p} \Erc(R,p,\Wch)$, where
\[
\Erc(R,p,\Wch) = \min_{\Vch} [ D(\Vch||\Wch|p) + |I(p,\Vch)-R|^+];
\]
see \cite{csiszar_koerner,CsiszarKoerner81a} for detail.
For consistency,
we assume all logarithms appearing in the definitions
of the rate of a code~\cite{gallager}
and functions $D$, $I$
are to base $\dmn$.
The function $E(R,P)$ coincides with $\Erc(R+1,p,\Wch)$
with $p$ being the uniform probability distribution on $\cX$
and $\Wch$ being the channel defined by $\Wch(v|u)=P(v-u)$, 
$u,v\in\cX=\field_{\dmn}\!\mbox{}^2$, where $\field_{\dmn}$ denotes
the finite field consisting of $\dmn$ elements.
Rewriting $\Erc(R,p,\Wch)$ into the other well-known form
(see \cite{csiszar_koerner}, pp.~168, 192--193, 
and \cite{gallager65,gallager}),
we have another form of $E$:
\[ 
E(R,P) = \max_{0\le \delta \le 1} -\delta (R-1) - (\delta+1) \log_{\dmn} \sum_{u\in\cX}
P(u)^{\frac{1}{\delta+1} }.
\]
Furthermore, putting
$\Pht_{\delta}(u)=P(u)^{\frac{1}{\delta+1}}/\sum_{v\in\cX}P(v)^{\frac{1}{\delta+1}}$,
$u\in\cX$, and $R_{\delta}=1-H(\Pht_{\delta})$, we obtain
\[
E(R,P) = \! \begin{cases}
-R+1 -2 \log_{\dmn} \sum_u   
P(u)^{\frac{1}{2}} 
& \! \mbox{if $0 \le R < R_1$},\\
D(\Pht_{\delta^{\star}}||P) & \!\!\!\!\! \mbox{if $R_1 \le R < R_0$},\\
\ \ \ \ \ 0 & \mbox{if $R_0\le R$},
\end{cases}
\]
where $\delta^{\star}$ is a $\delta$ with $R_{\delta}=R$;
see {\small FIG}.~\ref{fig:1}.

\section{Quantum Error-Correcting Codes} 

To prove the theorem, we use a lemma on codes for quantum channels.
We can regard the index of $\Ebe_{(i,j)}=X^i Z^j$, $(i,j)\in\cX$,
as a pair of elements from the field $\myF=\field_{\dmn}=\bZ/\dmn\bZ$.
From these, 
we obtain a basis $\Ebasis_n = \{ \Ebe_x \mid x \in (\myF^2)^n \}$
of $\Bop(\Hch^{\tnsr n})$,
where
$ 
\Ebe_x = \Ebe_{x_1} \tnsr \dots \tnsr \Ebe_{x_n}
$ 
for $x=(x_1,\dots,x_n)\in (\myF^2)^n$.
We write $\Ebe_{\Icr}$ for 
$\{ \Ebe_{x} \in \Ebasis_n \mid x\in\Icr \}$
where $\Icr \subset (\myF^2)^n$.
%
%
The index of a basis element
\[
((u_1,v_1),\dots,(u_n,v_n))\in (\myF^2)^n
\]
can be regarded as the plain $2n$-dimensional vector
\[
x=(u_1,v_1,\dots,u_n,v_n) \in \myF^{2n}.
\]
We can equip the vector space $\myF^{2n}$ over $\myF$ with
a {\em symplectic paring}\/ (bilinear form, or inner product) 
defined by
\[
\syp{x}{y} = \sum_{i=1}^{n} u_i v_i' - v_i u_i'
\]
for the above $x$ and $y=(u'_1,v'_1,\dots,u'_n,v'_n) \in \myF^{2n}$%
~\cite{artin,aschbacher}.
%
Given a subspace $\Cso \subset \myF^{2n}$, let
\[
\Cso^{\perp} = \{ x \in \myF^{2n} \mid \forall y\in \Cso,\ \syp{x}{y} =0 \}.
\]
\begin{lemma}\cite{crss97} 
\label{lem:symplectic_code} 
Let a subspace $\Cso\subset\myF^{2n}$ satisfy $\Cso\subset \Cso^{\perp}$
and $\dim \Cso = n-k$.
Choose a set $\Icr \subset \myF^{2n}$, not necessarily linear,
such that
\[
\{ y-x \mid x \in \Icr, y\in \Icr \} \subset (\Cso^{\perp} \setminus \Cso)\cmple,
\]
where the superscript ${\rm C}$ denotes complement.
Then, there exist $\dmn^k$-dimensional $\Ebe_{\Icr}$-correcting codes.
\end{lemma}
The codes in the lemma have the form
$ 
\{ \psi \in \Hch^{\tnsr n} \mid \forall M\in\Ebe_{\Cso},\ M \psi =  \tau(M) \psi \}
$ 
with some scalars $\tau(M)$, $M\in \Ebe_{\Cso}$.
A precise definition of $\Ebe_{\Icr}$-correcting codes
can be found in 
Sec.~III of \cite{KnillLaflamme97}
and the above lemma 
has been verified with Theorem III.2 therein.
Most constructions of quantum error-correcting codes relies on this lemma,
which 
is valid
even if $\dmn$ is a prime other than two%
~\cite{knill96a,knill96b,rains99,AshikhminKnill00}.

Now, for a memoryless channel
$\cA \sim \{ \sqrt{P(u)} \Ebe_{u} \}_{u\in\cX}$, 
and an $\Ebe_{\Icr}$-correcting code $\Hcd \subset \Hch^{\tnsr n}$,
write
\[
F(\Hcd) = \sup_{\cR_n} F(\Hcd,\cR_n \cA^{\tnsr n})
\]
where $\cR_n$ ranges over all TPCP linear maps on $\Bop(\Hch^{\tnsr n})$.
Then, 
since a recovery operator $\cR_n$ can be constructed explicitly so as to
correct all errors in $\Ebe_{\Icr}$, 
as in the proof of Theorem III.2 of \cite{KnillLaflamme97},
we have
\begin{equation}\label{eq:fid_bound}
1-F(\Hcd) \le \sum_{x\notin {\Icr}} P^{n}(x),
\end{equation}
where we have written $P^{n}(x_1\dots x_n)$ for
$P(x_1) \dots P(x_n)$.

\section{Proof of Theorem~1} 

%
We employ the method of types%
~\cite{csiszar_koerner,csiszar98,CsiszarKoerner81a},
on which a few basic facts to be used are 
collected here.
For $x=(x_1,\dots,x_n)\in\cX^n$,
define a probability distribution $\sP_{x}$
on $\cX$ by 
\[
\sP_{x}(u)=\crd{\{ i \mid 1\le i \le n, x_i = u \}}/n, \quad u \in \cX,	     
\]
which is called the {\em type}\/ (empirical distribution) of $x$. 
With $\cX$ fixed, the set of all possible types of sequences from
$\cX^n$ is denoted by $\cQ_n(\cX)$ or simply by $\cQ_n$.
For a type $Q\in \cQ_n$, $\cT_{Q}^n$ is defined as
$\{ x\in\cX^n \mid \sP_{x} = Q \}$.
In what follows, we use 
\begin{equation}\label{eq:types}
\crd{\cQ_n}\le (n+1)^{\crd{\cX}-1},\quad \mbox{and}\quad\forall Q\in \cQ_n,\
|\cT_{Q}^n| \le \dmn^{nH(Q)}.
\end{equation}
Note that if $x\in\cX^{n}$ has type $Q$, then
$P^{n}(x)=\prod_{a\in\cX} P(a)^{nQ(a)} = \exp_{\dmn} \{ -n [H(Q)+D(Q||P)]
\}$. 

We apply Lemma~\ref{lem:symplectic_code}
choosing $\Icr$ as follows.
Assume $\dim \Cso = n-k$.
Then, $\dim \Cso^{\perp} = n+k$~\cite{artin}. 
From each of the $\dmn^{n-k}$ cosets of $\Cso^{\perp}$ in $\myF^{2n}$,
select a vector that minimizes $H(\sP_{x})$, i.e., a vector $x$ satisfying
$H(\sP_{x})\le H(\sP_{y})$ for any $y$ in the coset.
This selection uses the idea of the minimum entropy 
decoder known in the classical information theory literature%
~\cite{CsiszarKoerner81a}.
Let $\Icr_0(\Cso)$ denote the set of the $\dmn^{n-k}$ selected vectors.
If we take $\Icr$ in Lemma~\ref{lem:symplectic_code} as 
$\Icr(\Cso)=\{ z+w \mid z\in\Icr_0(\Cso),w\in\Cso \}$,
the condition in the lemma is clearly satisfied.
Let
\[
\Aso = \{ \Cso \subset \myF^{2n} \mid \mbox{$\Cso$ linear}, \ \Cso \subset
\Cso^{\perp},\ \dim \Cso = n-k \},
\]
and  for each $L\in\Aso$, 
let $\Hcd(\Cso)$ be an $\Ebe_{\Icr(L)}$-correcting code
existence of which is ensured by 
Lemma~\ref{lem:symplectic_code}.
Put
\[
\Fbar = \frac{1}{\crd{\Aso}} \sum_{\Cso\in\Aso} F(\Hcd(\Cso)).
\]
We will show that $\Fbar$ is bounded from below by
$1 - (n+1)^{2(\dmn^2-1)} \dmn ^{ - n E(R,P) }$,
which establishes the theorem.
Such a method for a proof is called random coding%
~\cite{goppa74,csiszar_koerner,MatsumotoUyematsu01}.

The $\{0,1\}$-valued
indicator function 
$\indc [ T ]$ equals 1 if and only if the statement $T$ is true and
equals 0 otherwise.
From (\ref{eq:fid_bound}), we have
\begin{eqnarray}
1- \Fbar & \le & \frac{1}{\crd{\Aso}} \sum_{\Cso\in\Aso} \sum_{x \notin \Icr(\Cso)}  P^n(x) \notag\\
     & = & \frac{1}{\crd{\Aso}} \sum_{\Cso\in\Aso} \sum_{x \in \myF^{2n}}  
             P^n(x) \indc [ x \notin \Icr(\Cso) ]  \notag\\
     & = & \sum_{x \in \myF^{2n}} P^n(x) \frac{\crd{\Bcn(x)}}{\crd{\Aso}},
\label{eq:pr0}
\end{eqnarray}
where we have put 
\[
\Bcn(x) = \{ \Cso \in \Aso \mid x \notin \Icr(\Cso) \}, \quad x\in\myF^{2n}.
\]
The fraction $\crd{\Bcn(x)}/\crd{\Aso}$ is trivially bounded as
\begin{equation} \label{eq:pr1}
\frac{\crd{\Bcn(x)}}{\crd{\Aso}} \le 1, \quad x\in\myF^{2n}.
\end{equation}
We use the next inequality%
~\cite{crss97,MatsumotoUyematsu01}. 
Let
\[
\Acn(x) = \{ \Cso \in \Aso \mid x \in \Cso^{\perp}\setminus \Cso \}.
\]
Then, $\crd{\Acn(0)}=0$ and
\begin{equation}\label{eq:C_uniform}
\frac{\crd{\Acn(x)}}{\crd{\Aso}}
\le \frac{1}{\dmn^{n-k}}, \quad x\in\myF^{2n},\
 x \ne 0.
\end{equation}
%
Since $\Bcn(x) \subset \{ \Cso \in \Aso \mid \exists y\in\myF^{2n}, H(\sP_y) \le H(\sP_x), y-x\in \Cso^{\perp} \setminus \Cso \}$
from the design of $\Icr(\Cso)$ specified above (cf.~\cite{goppa74}), 
\begin{eqnarray}
 \crd{\Bcn(x)} &\le &\sum_{y\in \myF^{2n} :\, H(\sP_y) \le  H(\sP_x),\ y \ne x}
 \crd{\Acn(y-x)}\notag\\
  &\le & \sum_{y\in \myF^{2n} :\, H(\sP_y) \le  H(\sP_x),\ y \ne x}
\crd{\Aso}{\dmn}^{-n+k}, \label{eq:pr2} 
\end{eqnarray}
where we have used (\ref{eq:C_uniform}) for the latter inequality.
Combining (\ref{eq:pr0}), (\ref{eq:pr1}) and (\ref{eq:pr2}), we can proceed
as follows with the aid of the basic inequalities in (\ref{eq:types})
and the inequality
$\min \{ a+b, 1\} \le \min \{ a, 1\} + \min \{ b, 1\}$ for $a,b \ge 0$.
\begin{widetext}
\begin{eqnarray*}
1-\Fbar &\le& \sum_{x\in\myF^{2n}} P^n(x) \ \min \Biggl\{ \ \sum_{y\in\myF^{2n} :\, H(\sP_y) \le  H(\sP_x),\ y \ne x} \dmn^{-(n-k)},\ 1 \ \Biggr\}\\
 & \le & \sum_{Q\in \cQ_n}  \crd{\cT_{Q}^n} \prod_{a\in\cX} P(a)^{nQ(a)}
\ \min \Biggr\{ \sum_{Q'\in \cQ_n :\, H(Q') \le H(Q)} \frac{|\cT_{Q'}^n|}{\dmn^{n(1-R)}}, \ 1\ \Biggl\}\\
& \le & \sum_{Q\in\cQ_n} \exp_{\dmn} [ -n D(Q || P) ] 
\sum_{Q'\in \cQ_n :\, H(Q')\le H(Q)} \exp_{\dmn} [ -n |1-R-H(Q')|^{+} ]\\
& \le & \sum_{Q\in\cQ_n} \exp_{\dmn} [ -n D(Q || P) ] \,
|\cQ_n| \max_{Q'\in\cQ_n :\, H(Q') \le H(Q)}\exp_{\dmn} [ -n |1-R-H(Q')|^{+} ]\\
& = & \sum_{Q\in\cQ_n} |\cQ_n| \exp_{\dmn} [ -n D(Q||P) -n|1-R-H(Q)|^+ ]\\
&\le & (n+1)^{2(\dmn^2-1)} \exp_{\dmn} [ -n E(R,P) ],
\end{eqnarray*}
\end{widetext}
which is the promised bound. 

\section{Concluding Remark}

This author conjectures that the bound in (\ref{eq:new_exp}) 
is not tight in view
of the existence of the Shor-Smolin codes~\cite{dss98}.

\section*{Acknowledgments}

The author 
would like to thank R.~Matsumoto for 
valuable information on algebraic matters, discussions, comments,
and especially, pointing out an error 
in the earlier manuscript, 
M.~Hayashi and K.~Matsumoto
for helpful comments,
and H.~Imai for support.

\end{document}